\documentclass[10pt,amsmath,aps,amssymb,superscriptaddress,notitlepage,onecolumn,longbibliography]{revtex4-1}

\makeatletter
\renewcommand*{\p@subsubsection}{}
\makeatother

\usepackage{bm,amsmath,amssymb,
}
\usepackage{natbib} 
\usepackage{amsmath}
\usepackage{graphicx}
\usepackage{mathtools}
\usepackage{mathtools}
\usepackage{float}
\usepackage{empheq}
\usepackage{epstopdf}
\usepackage{epstopdf}
\usepackage{amssymb}
\usepackage{wrapfig}
\usepackage{picture}
\usepackage{mwe,tikz}
\usepackage{color}
\usepackage{transparent}
\usepackage{textcomp}
\makeatletter
\newcommand{\verbatimfont}[1]{\def\verbatim@font{#1}}%
\makeatother

\definecolor{Lblu}{RGB}{193,193,234}
\definecolor{Mblu}{RGB}{131,131, 221}
\definecolor{Dblu}{RGB}{0,0, 255}
\definecolor{Lgr}{RGB}{192,192,192}
\definecolor{Mgr}{RGB}{138,138,138}

\definecolor{OR}{RGB}{244,127,31} 
\definecolor{BR}{RGB}{170,0,0} 
\definecolor{GR}{RGB}{26,211,26} 
\definecolor{VIO}{RGB}{170,0,170} 
\definecolor{PI}{RGB}{237,159,165} 

    \def\XXint#1#2#3{{\setbox0=\hbox{$#1{#2#3}{\int}$}
         \vcenter{\hbox{$#2#3$}}\kern-.5\wd0}}

\newcommand{\tensbf}[1]{\boldsymbol{\mathsf #1}} 

\newcommand{\bff}{{\mathbf f}}

\newcommand{\bu}{{\mathbf u}}

\newcommand{\bx}{{\mathbf x}}

\newcommand{\md}{{\mathrm{d}}}

\begin{document}
	
\title{Trapping, gliding, vaulting: Transport of semiflexible polymers in periodic post arrays}
\author{Brato Chakrabarti}
\affiliation{Department of Mechanical and Aerospace Engineering, University of California San Diego,
	9500 Gilman Drive, La Jolla, CA 92093, USA}
\affiliation{~Center for Computational Biology, Flatiron Institute, 162 5th Ave, New York, NY 10010.}
\author{Charles Gaillard}
\affiliation{\'Ecole Polytechnique, Route de Saclay, 91128 Palaiseau, France}
\author{David Saintillan}
\affiliation{Department of Mechanical and Aerospace Engineering, University of California San Diego,
	9500 Gilman Drive, La Jolla, CA 92093, USA}

\date{\today}

\begin{abstract}
The transport of deformable particles through porous media underlies a wealth of applications  ranging from filtration to oil recovery to the transport and spreading of biological agents. Using direct numerical simulations, we analyze the dynamics of semiflexible polymers under the influence of an imposed flow in a structured two-dimensional lattice serving as an idealization of a porous medium. This problem has received much attention in the limit of reptation and for long-chain polymer molecules such as DNA that are transported through micropost arrays for electrophoretic chromatographic separation. In contrast to long entropic molecules, the dynamics of elastic polymers  results from a combination of scattering with the obstacles and flow-induced buckling instabilities. We identify three dominant modes of transport that involve trapping, gliding and vaulting of the polymers around the obstacles, and we reveal their essential features using tools from dynamical systems theory. The interplay of these scattering dynamics with transport and deformations in the imposed flow results in the long-time asymptotic dispersion of the center of mass, which we quantify in terms of a hydrodynamic dispersion tensor. We then discuss a simple yet efficient chromatographic device that exploits the competition between different modes of transport to sort filaments in a dilute suspension according to their lengths.
\end{abstract}

\maketitle
\section{Introduction}
The transport of clouds of particles through complex structured media underlies a variety of important physical processes in both nature and industry. Examples range from the spreading of contaminants in porous media \cite{edwards1991dispersion} to solute transport in biofilms \cite{dykaar1996macrotransport} to the dispersion of engineered drugs inside tumors \cite{felfoul2016magneto} to membrane filtration processes. These problems often involve the spreading of an initially concentrated collection of particles as they are transported through the tortuous geometry under the action of an external flow or force and in the presence of molecular diffusion. For a large number of such problems, the long-time transport process can be described by a mean velocity $\mathbf{U}$ and effective hydrodynamic dispersion tensor $\tensbf{D}$ that depend on the \emph{microtransport} dictated by flow topologies, physiochemical processes and geometry of the microstructure. The theoretical description of these asymptotic transport coefficients forms the basis of \emph{macrotransport} theory \cite{brenner2013macrotransport} that serves as the backbone of many industrial applications ranging from filtration to the design of chromatographic devices.

Even though the macrotransport theory of point-like passive \cite{brenner1980dispersion} and active \cite{alonso2019transport} Brownian particles in porous media is well developed, modeling the transport of elongated or deformable finite-size particles remains a challenge due to non-trivial particle-obstacle interactions and excluded volume effects. Perhaps one of the earliest examples where this problem shows up is in the celebrated reptation theory of de Gennes \cite{de1971reptation}, which is concerned with the thermal motion of a long linear polymer chain past fixed obstacles serving as a model of entangled macromolecules in polymer melts \cite{doi1988theory}. The diffusion of stiffer semiflexible filaments in porous media has also been shown to follow the reptation picture, albeit with different kinetic exponents \cite{nam2010reptation}. Understanding the transport of flexible polymers in porous media under the application of external forces also has extensive applications in chromatographic device designs for long-chain DNA molecules \cite{patel2003computational,mohan2007stochastic}. Fast and efficient size-dependent separation of DNA molecules plays an important role in their mapping and sequencing, both crucial for genomic analysis \cite{dorfman2012beyond}. Compared to classical gel-based electrophoretic separation, modern microfluidic chromatographic devices have  proven to be much more efficient for these problems \cite{sambrook1989molecular}. In these devices, DNA molecules are transported in a 2D lattice of structured microposts under the application of an external electric field. The DNA repetitively collides with the posts of the array, with a size-dependent collision time leading to rapid separation \cite{chou1999sorting}.  DNA molecules  have a persistence length $\ell_p$ that is much smaller than their contour length $L$, with dynamics governed by a competition between stretching and entropic preference of a coiled state. During transport, the molecules hook and unhook from the microposts with dynamics similar to a rope over pulley and conformations resembling various English alphabets, which have been studied extensively in experiments, simulations and through continuous time random walk models \cite{chou1999sorting,sevick2001long,kim2007brownian,teclemariam2007dynamics,cho2010brownian,dorfman2010dna,olson2011continuous,kawale2017polymer}.

Contrary to long-chain polymer molecules, the dynamics of semiflexible polymers with  $L \sim \ell_p$ is dominated by a competition between local bending forces, line tension that enforces inextensibility, and thermal fluctuations. These make way for a number of buckling instabilities and lead to non-trivial filament conformations that have been well characterized in unbounded flows \cite{kantsler2012fluctuations,manikantan2015buckling,liu2018morphological,chakrabarti2019flexible}. Filament transport has also been studied in cellular flows where an instability-driven stretch-coil transition can lead to diffusive or sub-diffusive transport of the center-of-mass \cite{young2007stretch,manikantan2013subdiffusive,quennouz2015transport}. However, the dynamics of stiff polymers in structured porous media has received little attention, with most studies restricted to the limit of reptation \cite{nam2010reptation,milchev2011single}. Understanding their transport in crowded environments is relevant for stiff biopolymers such as actin and microtubules \cite{mokhtari2019dynamics}, for the locomotion of micro-organisms through granular materials \cite{majmudar2012experiments} and for biological agents having the potential to maximize their transport through interactions with their local environment \cite{ohta2009deformable}.

In this paper, we use direct numerical simulations to study the transport of semiflexible polymers modeled as fluctuating inextensible Euler elastica through a two-dimensional periodic lattice of circular obstacles under a pressure-driven flow. We aim to characterize the essential features of transport, which results from the coupling between deformations due to dynamic buckling and polymer-obstacle scattering, leading to a diffusive behavior at long times. This is in contrast to active filaments that exhibit sub-diffusive transport in disordered media as shown recently in numerical simulations \cite{mokhtari2019dynamics}.  In Section \ref{sec:sec2}, we discuss the theoretical model for the fluctuating polymer and its dynamics in the lattice. In Section \ref{sec:sec3}, we identify three main modes of polymer transport and describe how they can be used to explain asymptotic hydrodynamic dispersion. In Section \ref{sec:sec4}, we build on our understanding of polymer-obstacle interactions to propose a simple design of a chromatographic device that is able to sort polymers based on their lengths. We conclude in Section \ref{sec:sec5}.

\section{Problem formulation and methods}\label{sec:sec2}

\subsection{Polymer and lattice models}

\begin{figure}[b]
	\centering
	\includegraphics[width=0.45\linewidth]{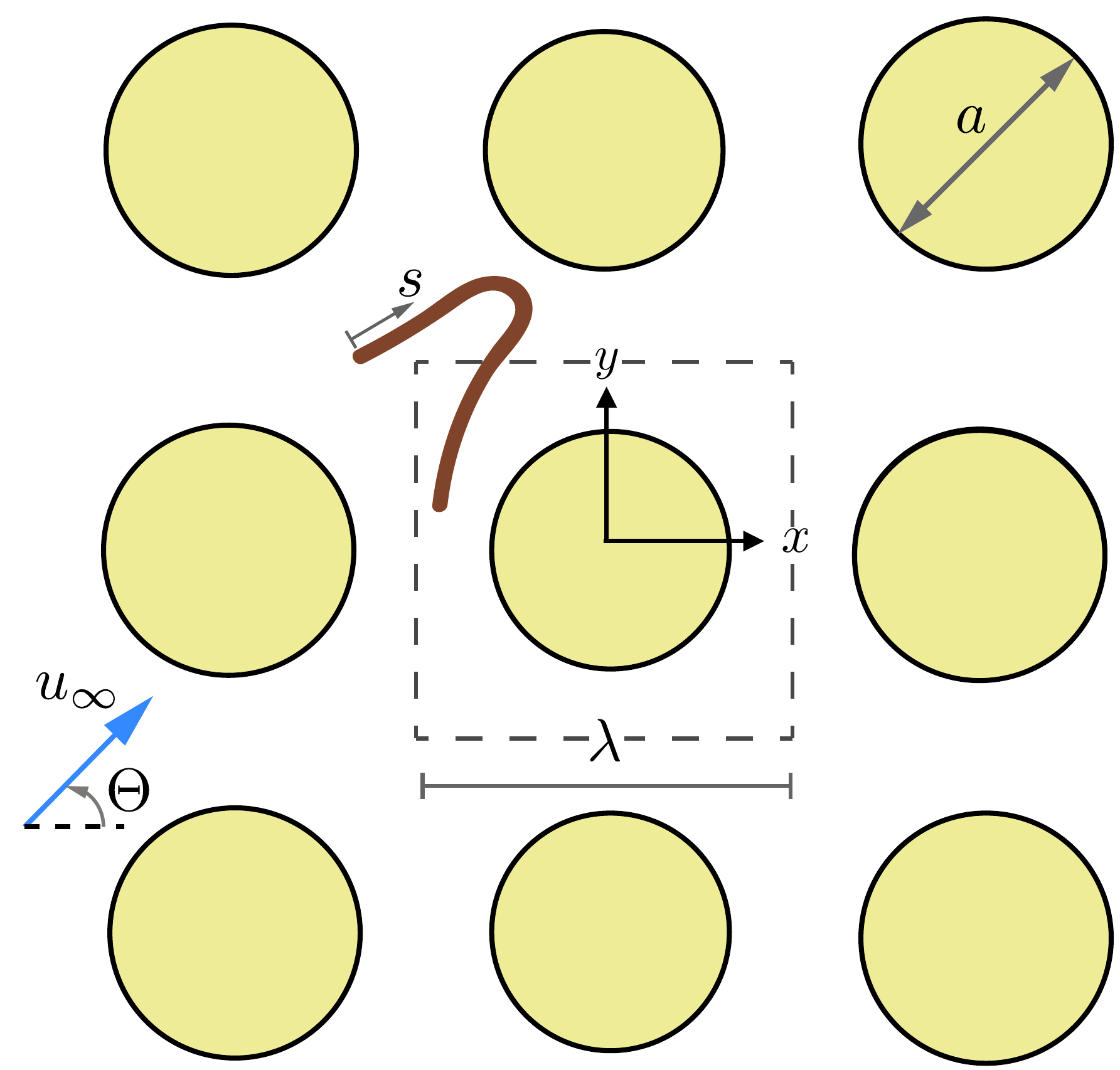}
	\caption{Schematic of the 2D porous lattice and a representative unit cell.}
	\label{fig:lattice}
\end{figure}

We study the dynamics, conformations and long-time asymptotic transport of semi-flexible polymers with $L \sim \ell_p$ through a doubly periodic two-dimensional porous medium under the influence of an imposed flow. The porous medium, as shown in Fig.~\ref{fig:lattice}, is idealized as an infinite lattice comprised of rigid circular obstacles of diameter $a$. The distance between successive pillars $\lambda$ is identical in the $x$ and $y$ directions resulting in a representative square unit cell. The ordered array is characterized by its porosity $\epsilon = S_f/S_t$, or ratio of the  fluid area $S_f$ of a unit cell over its total area $S_t$. 
The semiflexible filaments are modeled as inextensible slender elastic rods of length $L$ with circular cross-section of diameter $d$.  The centerline of the polymer is represented as a space curve parameterized by arclength $s$ and identified by a Lagrangian marker $\bx(s,t)$. 
Filament dynamics are modeled using local slender body theory \cite{keller1976slender,tornberg2004simulating,du2019dynamics} as
\begin{equation}
8 \pi \mu \left(\frac{\partial \bx(s,t)}{\partial t} - \bu(\bx(s,t))\right) = -\tensbf{\Lambda} \cdot \bff,
\end{equation}
where $\mu$ denotes the viscosity of the fluid, $\mathbf{u}$ is the velocity of the imposed flow, $\tensbf{\Lambda}$ is the local mobility operator, and $\bff$ is the force per unit length exerted by the filament on the fluid. We note that this leading-order anisotropic drag model neglects inter-filament and filament-obstacle hydrodynamic interactions; we discuss this approximation further in Section~\ref{sec:sec5}. The configuration-dependent local operator is given by:
\begin{equation}
\tensbf{\Lambda}(s) = (2-c)\mathbf{I} - (c+2) \bx_s \bx_s,
\end{equation}
where subscript $s$ denotes differentiation with respect to arc-length and $\bx_s$ is the local tangent vector. $c$ is an asymptotic geometric parameter that depends on the slenderness of the filament and is defined as $c = - \ln(\alpha^2 e)$ where $\alpha = d/L$.  For the chosen model of elasticity, the force density is given by:
\begin{equation}
\bff  = B \bx_{ssss} - (\sigma \bx_s)_s + \bff^{Br},
\end{equation}
where the first term is typical of Euler-Bernoulli beam theory with bending rigidity $B$. The inextensibility of the filament results in a metric constraint $\bx_s \cdot \bx_s = 1$, which gives rise to the second term where $\sigma$ acts as a Lagrange multiplier and can be interpreted as the internal line tension. The third term accounts for Brownian fluctuations and obeys the fluctuation-dissipation theorem:
\begin{align}
\langle \bff^{Br}(s,t) \rangle &= \mathbf{0}, \\
\langle \bff^{Br}(s,t)\bff^{Br}(s',t') \rangle &= 2 k_B T \tensbf{\Lambda}^{-1} \delta(t-t')\delta(s-s'),
\end{align}
where $k_B$ is the Boltzmann constant and $T$ is temperature. The persistence length $\ell_p = B/k_B T$ characterizes the distance along the centerline over which the unit tangent vector  loses correlation with itself.

\begin{figure}[t]
	\centering
	\includegraphics[width=0.7\linewidth]{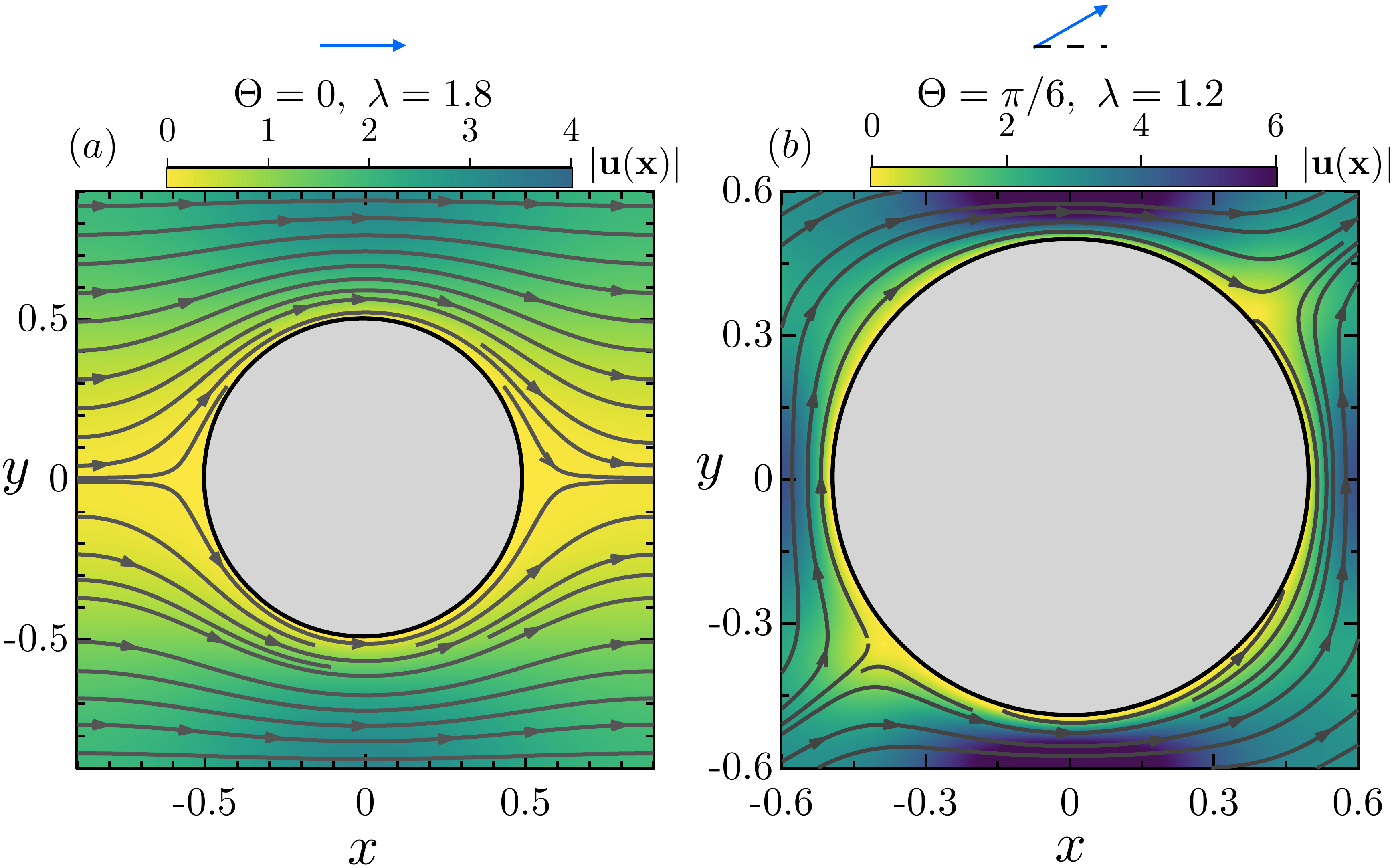}
	\caption{Flow streamlines inside a unit cell for two different lattice porosities and mean incidence angles of the applied flow.}
	\label{fig:stream}
\end{figure}

The imposed fluid velocity $\bu(\bx)$ is taken to be that induced by a macroscopic pressure gradient applied across the array, with a far-field velocity of $u_\infty$. The velocity field inside the unit cell is obtained numerically by solving the Stokes equations using the  boundary integral method with an appropriate choice of Green's function \cite{alonso2019transport,pozrikidis1992boundary}.   Computed streamlines for two representative cases are shown in Fig.~\ref{fig:stream}, where the flow topologies are found to be governed by the distance between obstacles $\lambda$ and the incidence angle $\Theta$ made by the applied flow with respect to the $x$ direction.

In the following, we non-dimensionalize all the equations using the diameter $a$ of the pillars as the characteristic length scale, $u_\infty$ as the velocity scale for the applied flow, $B/L^2$ as the scale for elastic forces, $\sqrt{L/\ell_p}B/L^2$ as the scale for Brownian forces \cite{manikantan2013subdiffusive}, and the relaxation time of the polymer $\tau = 8 \pi \mu L^4/Bc$ as time scale. With these choices, the dimensionless governing equation is given by:
\begin{equation}\label{eq:sbt}
\frac{\partial \bx(s,t)}{\partial t} = \bar{\mu} \bu(\bx(s,t)) - \textcolor{black}{\frac{L}{a}}\tensbf{\Lambda} \cdot \Big(\bx_{ssss} - (\sigma \bx_s)_s + \sqrt{L/\ell_p}\bm{\xi}\Big).
\end{equation}
Two important dimensionless numbers appear.\ The elastoviscous number  $\bar{\mu} = 8 \pi \mu L^4 u_\infty/B a c$ compares the time scale of bending relaxation to the characteristic inverse shear rate $a/u_\infty$, and serves as an effective measure of the strength of the applied flow. The ratio $\ell_p/L$ captures the magnitude of thermal fluctuations, with the limit of $\ell_p/L\rightarrow \infty$ describing Brownian rigid rods. $\bm{\xi}$ is a Gaussian random vector with zero-mean and unit variance. As mentioned previously, the resulting dynamics depend strongly on the flow topology and geometry of the microstructure. This is characterized by three additional dimensionless parameters:
\begin{equation}
\frac{L}{a}, \hspace*{4mm} \epsilon = 1-\frac{\pi a^2}{\lambda^2} \hspace*{2mm} \text{  and  } \hspace*{2mm} \Theta,
\end{equation}
where $L/a$ compares the filament length to the obstacle diameter, $\epsilon$ is the porosity and $\Theta$ is the mean direction of the applied flow with respect to the $x$ axis. In all the results presented in the paper, we set $\ell_p/L=20$, and focus on the effects of flow strength and lattice geometry. We also restrict our results to $L/a=0.7$ until the discussion of polymer sorting in Section \ref{sec:sec4}, where the effect of contour length is examined.

\subsection{Numerical method}

Associated with eqn \eqref{eq:sbt} are force- and moment-free boundary conditions that translate to $\sigma = \bx_{ss} = \bx_{sss} = 0$ at $s = 0, \ L/a$. We exploit the inextensibility condition to solve for the unknown line tension $\sigma$ and use an implicit-explicit time-marching scheme to solve \eqref{eq:sbt} with the appropriate boundary conditions; details of the methods can be found elsewhere\cite{tornberg2004simulating,liu2018morphological}. \textcolor{black}{The typical integration time step was in the range of $\Delta t = 2 \times 10^{-9}-10^{-7}$ in dimensionless units. Our computational model  for the polymer was validated by comparing it to well-known equilibrium properties of semiflexible polymers \cite{milchev2018nematic}, details of which can be found in \cite{manikantan2015bending}. To study asymptotic transport properties, we averaged over 150 filament trajectories over a distance of 700 unit cells.} In order to facilitate fast computation, the interstitial velocity field $\bu(\bx)$ was tabulated on a Cartesian mesh, and bilinear interpolation was used subsequently. \textcolor{black}{Due to the symmetry of the lattice geometry and linearity of Stokes flow, we computed the velocity in only one quadrant of the unit cell on a Cartesian grid of 250$\times$250 mesh points.}

Central to the present study is the prescribed mechanism of polymer-pillar scattering. For this, we allow the filaments to have tangential motion past the obstacles, resulting in a \emph{gliding} behavior. Motions into the pillars are avoided by a smooth hydrodynamic repulsion force as first proposed by Evans \textit{et al.}\cite{evans2013elastocapillary}, which efficiently prevents filament penetration. Details of the contact algorithm are described in the Appendix.

\begin{figure}
	\centering
	\includegraphics[width=0.65\linewidth]{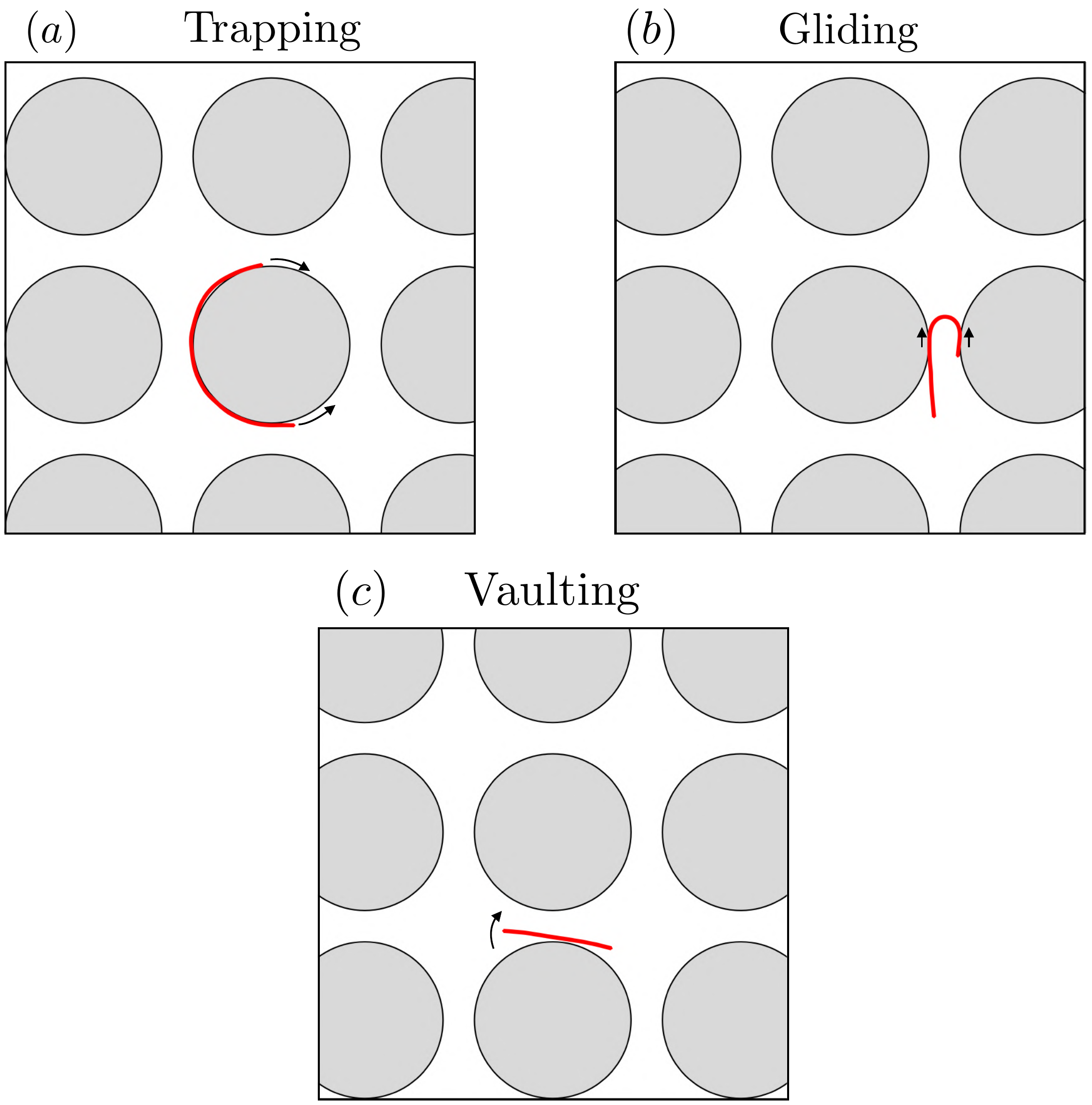}
	\caption{Typical filament conformations and modes of transport: ($a$) trapping, ($b$) gliding, and ($c$) vaulting. The arrows indicate possible directions of motion.}
	\label{fig:trans}
\end{figure}

\begin{figure}
	\centering
	\includegraphics[width=0.95\linewidth]{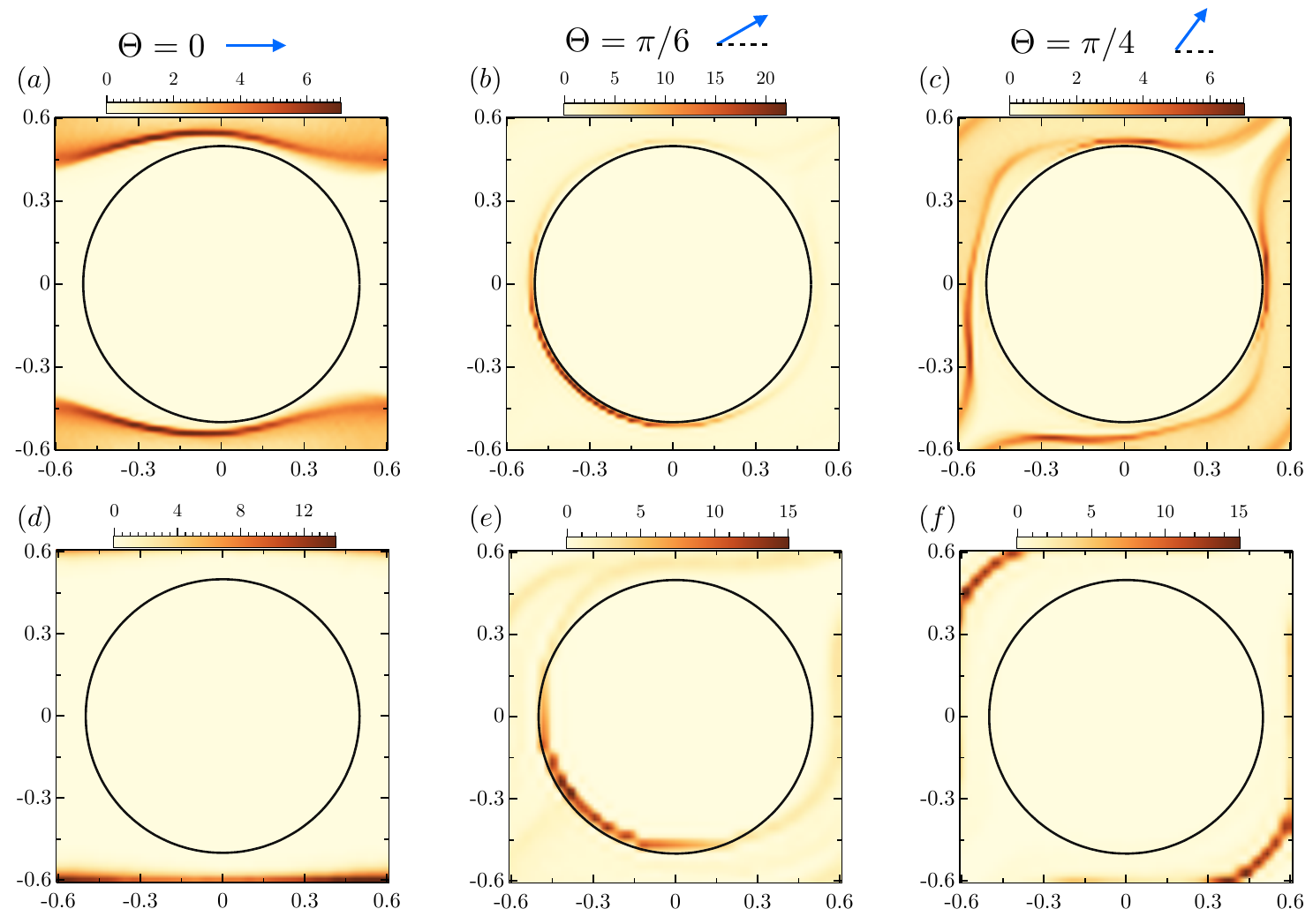}
	\caption{($a$)--($c$) Probability distribution of the polymer chain for different incidence angles $\Theta$ of the applied flow. ($d$)--\textcolor{black}{($f$)} Probability distribution of the polymer center of mass for the same angles. The black circle represents the circular pillar. Parameter values: $L/a = 0.7$, $\epsilon = 0.45$ and $\bar{\mu} = 8 \times 10^3$.}
	\label{fig:angdist}
\end{figure}

\section{Results and discussion}\label{sec:sec3}

\subsection{Pore-scale dynamics and modes of transport}

We begin with a qualitative discussion of the key features of individual filament dynamics at the pore scale, which are expected to dictate long-time transport properties. The dynamics of elastic filaments in unbounded linear flows were characterized in detail in our previous work  \cite{liu2018morphological,chakrabarti2019flexible} and provide a basis with which to compare dynamics in the present problem. In a weak shear flow, a Brownian fiber typically tumbles quasi-periodically in what is reminiscent of the classical Jeffery orbits \cite{jeffery1922motion}. Increasing the elastoviscous number $\bar{\mu}$ triggers a sequence of  buckling instabilities that result in transient $C$ and $U$ shaped conformations. In a porous lattice, shear is generated between pillars and can also cause tumbling and buckling, though velocity gradients are typically non-uniform on the scale of the filament. In addition, scattering dynamics through the lattice allow for more complex motions, where a filament can now deform and buckle while interacting with obstacles even in relatively weak flows. In the explored parameter space, we identified three dominant modes of transport, which we illustrate in Fig.~\ref{fig:trans} and Supplementary Videos and whose qualitative features are as follows:

\begin{itemize}
	\item \textit{Trapping:} During transport, a filament can wrap around the circular obstacles as shown in Fig.~\ref{fig:trans}($a$), resulting in periods of prolonged trapping. As expected, these trapping events are more frequent for long polymers or in strong flows through densely packed lattices. When a filament gets trapped, it then relies on Brownian motion for shape fluctuations,  which allow it to escape by sliding in either direction along the obstacle.
	
	\item \textit{Gliding and squeezing:} Irrespective of the filament length and porosity of the lattice, a polymer can glance or slide past obstacles. In dense beds, this results in a \emph{squeezing} mode as illustrated in Fig.~\ref{fig:trans}($b$). The incidence angle $\Theta$ of the incoming flow plays an important role in the selection between trapping and gliding. For $\Theta=0$ or $\pi/2$ the filaments almost always avoid being trapped and can get locked in the {squeezing} mode of transport with quasi-periodic dynamics as we reveal in the subsequent discussions.
	
	\item \textit{Vaulting:} Finally, below the threshold of buckling, we observe Brownian vaulting of the filaments. These vaults, as shown in Fig.~\ref{fig:trans}($c$), are distinct from \textcolor{black}{Jeffery} orbits in unbounded flows. In this mode, a filament uses its contact point with a pillar as a hinge to slide, rotate and move forward in a fashion similar to what is observed during the sedimentation of fibers near walls \cite{russel1977} or the transport of fibers in strongly confined microchannels \cite{nagel2018oscillations}.
\end{itemize}
Note that in most cases a combination of multiple modes of transport is observed along filament trajectories, with alternating trapping, gliding and vaulting events. As we explain next, the relative proportion of various modes strongly depends on incoming flow angle as well as lattice geometry and flow strength, and can be exploited for filament separation as discussed in Section~\ref{sec:sec4}.

\setcounter{figure}{5}
\begin{figure}[!b]
	\centering\vspace{0.15cm}
	\includegraphics[width=0.95\linewidth]{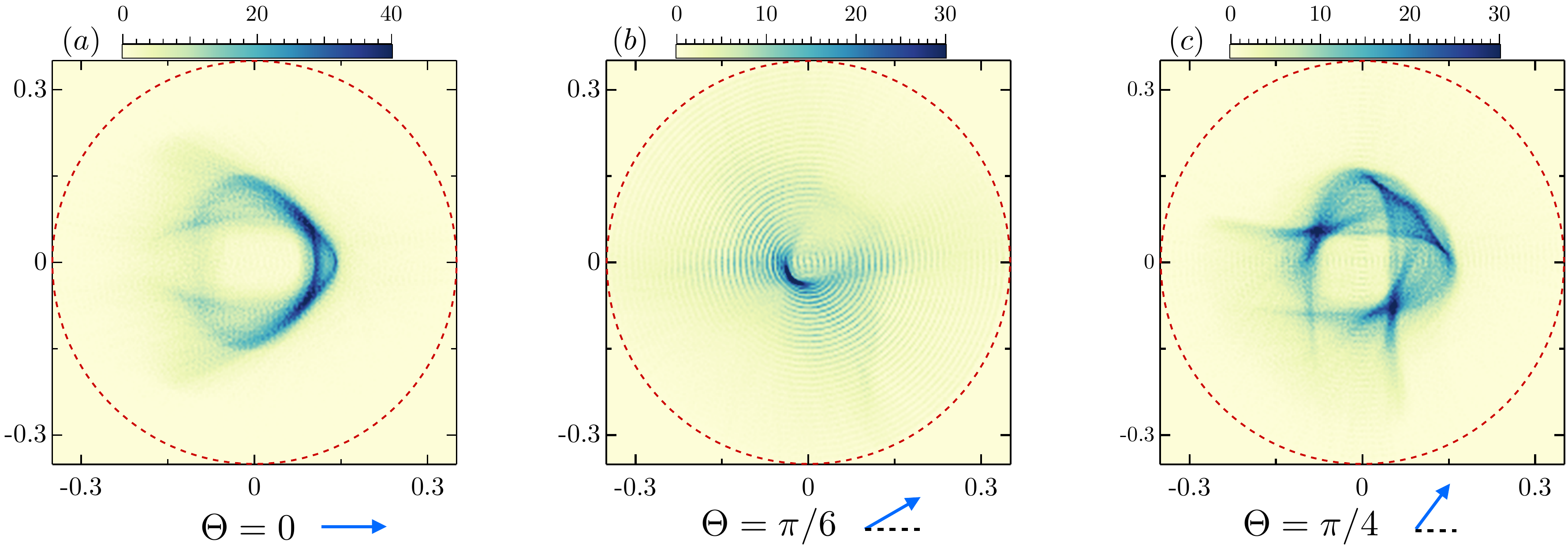}\vspace{-0.15cm}
	\caption{($a$)-($c$) Probability distribution of overlayed filament conformations with their COM at the origin, for different incidence angles $\Theta$ of the flow. The red circle with diameter $L/a=0.7$ represents the allowable spread of the filament. Parameter values are as in Fig.~\ref{fig:angdist}.}
	\label{fig:strob}
\end{figure}

\subsection{Probability distributions and filament trajectories}

With our understanding of the three dominant modes of transport, we now proceed to discuss the main features of filament trajectories as a function of the various dimensionless numbers.  We first consider the probability distribution function of the entire polymer chain inside a representative unit cell. The distribution is computed by averaging over all the unit cells visited by the polymer and is subsequently normalized to unity. Fig.~\ref{fig:angdist}($a$)-($c$) shows this distribution in a dense lattice for different incidence angles $\Theta$ of the applied flow. We notice that for $\Theta=0$ and $\pi/4$ this distribution has a mirror symmetry with respect to the flow direction. As discussed previously and also evident from Fig.~\ref{fig:angdist}($a$), we find that for $\Theta=0$ the filament is locked in a \emph{squeezing} mode between two rows of pillars and primarily relies on gliding between the obstacles to move forward. This is further highlighted in Fig.~\ref{fig:angdist}($d$) showing the associated center-of-mass (COM) distribution of the polymer. The COM distribution peaks at the symmetry plane of the lattice and is vanishingly small elsewhere inside the unit cell. This further corroborates the caged dynamics of the polymer and underscores its inability to diffuse in the vertical direction with the spaces between pillars acting as entropic traps \cite{han2000separation}. \emph{Gliding} still remains the primary mode of transport for $\Theta=\pi/4$, however the dynamics is not caged in this case. Due to symmetry, the filament can glide either in the $x$ or $y$ direction with equal probability, resulting in the distribution shown in Fig.~\ref{fig:angdist}($c$) that bears resemblance with the streamlines of the flow.

A transition to the \emph{trapping} mode is observed when the incidence angle is $\Theta=\pi/6$. The lack of symmetry in this case results in chaotic scattering dynamics with frequent trapping events around the obstacles. This is evident in Fig.~\ref{fig:angdist}($b$) where the distribution peaks only in the vicinity of the pillar. The accompanying COM distribution in Fig.~\ref{fig:angdist}($e$) shows a peak inside the obstacle, corresponding to configurations where the polymer is wrapped around the pillar, a characteristic feature of the trapping mode as shown in Fig.~\ref{fig:trans}($a$). A similar transition from \emph{gliding} to \emph{trapping} also takes place as a function of  flow strength. This is illustrated in Fig.~\ref{fig:fldist}, where we show the COM probability distribution for two different values of $\bar{\mu}$ with $\Theta=\pi/6$. While in weak flow both trapping and gliding contribute to transport, we find that the polymer remains predominantly trapped in stronger flows. The dominant mode of transport in this case is selected from a competition between flow-induced buckling instabilities favoring deformed conformations and sliding resulting from filament-obstacle interactions.

\setcounter{figure}{4}
\begin{figure}[t]
	\centering
	\includegraphics[width=0.7\linewidth]{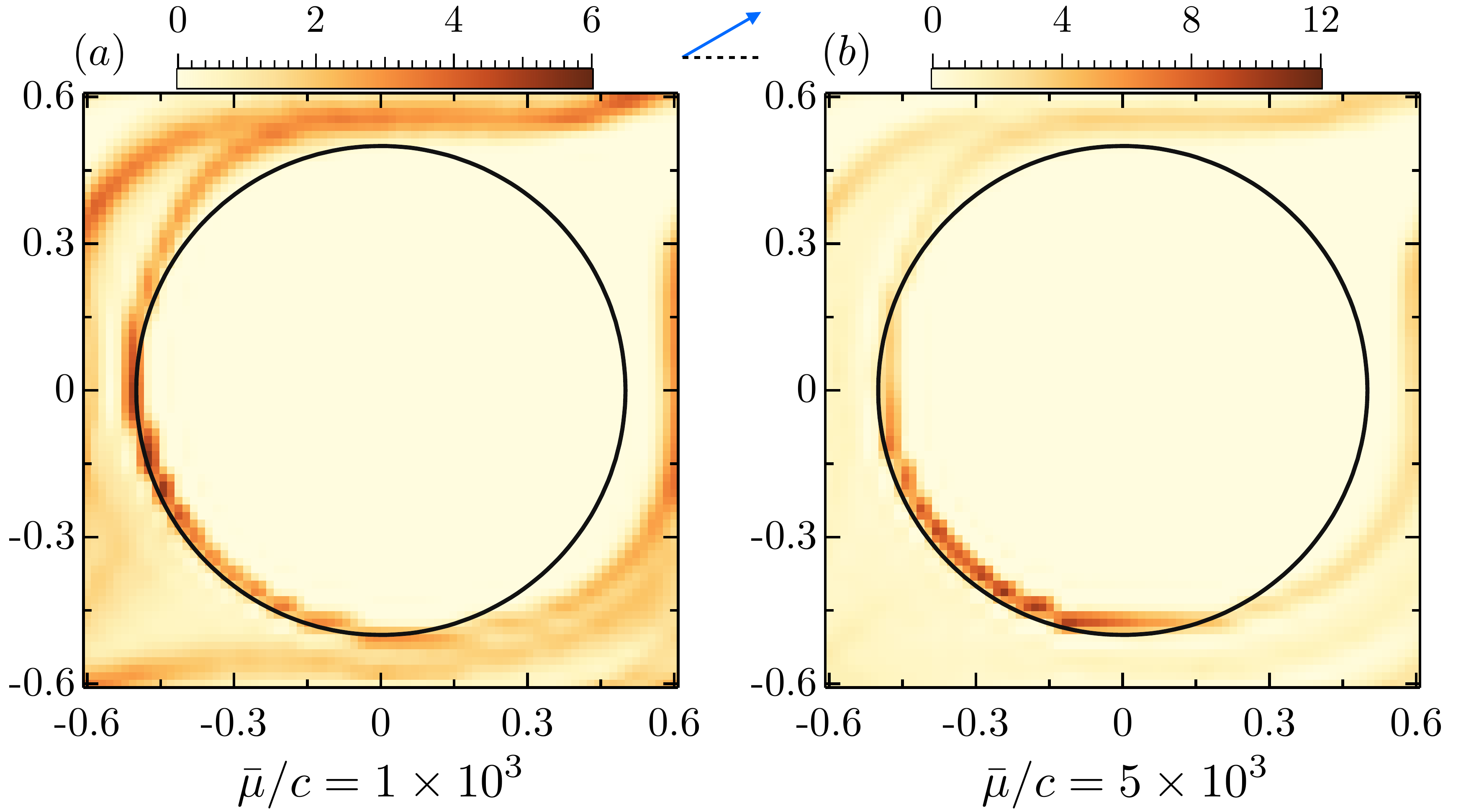}
	\caption{Probability distribution function of the polymer COM for two different flow strengths $\bar{\mu}$. Parameter values are as in Fig.~\ref{fig:angdist} with $\Theta = \pi/6$.}\vspace{-0.2cm}
	\label{fig:fldist}
\end{figure}

These features of the scattering process are further revealed in Fig.~\ref{fig:strob}, where we display distributions of successive filament conformations overlaid by subtracting the instantaneous COM. Fig.~\ref{fig:strob}($a$) now clearly captures the caged behavior of the filament, where we observe two dominant conformations, both resembling the rotated letter $U$, one more concave than the other. As the filament squeezes between two rows of pillars, it oscillates between these two dominant shapes in a breathing pattern. Conformations for angle $\Theta=\pi/4$ exhibit a sweeping pattern spanning an angle of $\pi/2$ that results from symmetric gliding in the horizontal and vertical directions. The chaotic scattering process for $\Theta=\pi/6$ is captured in Fig.~\ref{fig:strob}($b$), where we observe a zoo of conformations with no distinct peak, hinting at the randomness of the process.

To further characterize these dynamics, we introduce the gyration tensor of the filament defined as: 
\begin{equation}
G_{i j}(t)=\frac{1}{L} \int_{0}^{L}\left[x_{i}(s, t)-\bar{x}_{i}(t)\right]\left[x_{j}(s, t)-\bar{x}_{j}(t)\right] \,d s,
\end{equation}
where $\bar{\bx}(t)$ is the filament COM. The
angle $\phi$ between the mean filament orientation and the flow
direction is provided by the eigenvector of $G_{ij}$ associated with the dominant eigenvalue \cite{liu2018morphological}.
Fig.~\ref{fig:PSD} shows the power spectral density of $\phi(t)$ as a function of dimensionless frequency. \textcolor{black}{We first vary the incoming flow angle in Fig.~\ref{fig:PSD}$(a)$.} For $\Theta=0$, we notice that there are two sharp peaks in the power spectral density that can be mapped back to the filament breathing between two dominant conformations in an almost time-periodic trajectory. For $\Theta=\pi/4$, we observe multiple peaks in the spectrum indicative of quasi-periodicity \cite{strogatz2018nonlinear} that results from gliding of several repeating conformations in the lattice. Both $\Theta=\pi/6$  and $\Theta = \pi/8$ are characterized by a broad power spectrum with no clear peaks. \textcolor{black}{This is a signature of chaotic trajectories that result from trapping of filaments around obstacles with a large distribution of stopping times. In Fig.~\ref{fig:MSD}$(b)$, we keep the angle of the incoming flow fixed at $\Theta = \pi/4$ but vary the length of the polymer. The short filament does not deform and performs vaulting around the obstacles, which results in two peaks in the power spectrum. Surprisingly the longer filament, while exhibiting a broad power spectrum, performs a distinct periodic motion with a well-defined peak. In this regime, we find that the filament adopts coiled conformations that are preserved as the filament gets transported through the lattice. Finally, for the intermediate length we observe the quasiperiodic trajectories already described above.}

\setcounter{figure}{6}
\begin{figure}[H]
	\centering
	\includegraphics[width=1\linewidth]{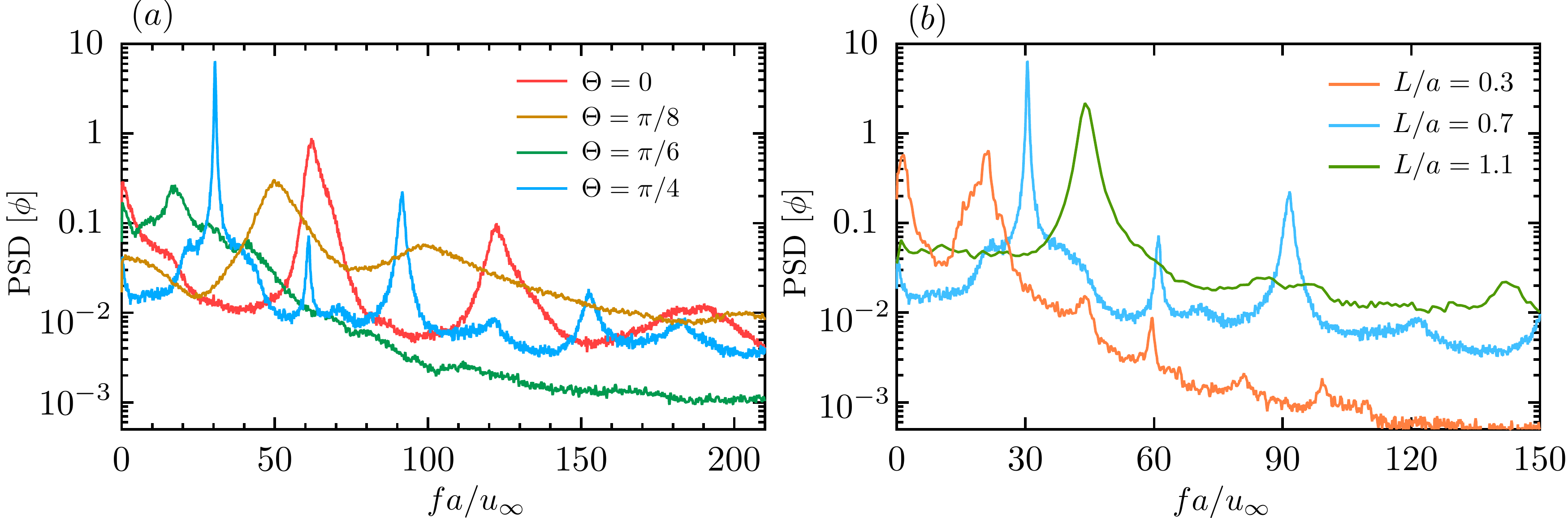}\vspace{-0.1cm}
	\caption{Power spectral density of the mean orientation angle $\phi(t)$ of the filament with respect to the flow direction as a function of dimensionless frequency, for different flow orientations ($a$) and lengths of the polymer ($b$). Parameter values are as in Fig.~\ref{fig:angdist}.}\vspace{-0.0cm}
	\label{fig:PSD}
\end{figure}

\subsection{Asymptotic transport and hydrodynamic dispersion}

\begin{figure}[t]
	\centering
	\includegraphics[width=1\linewidth]{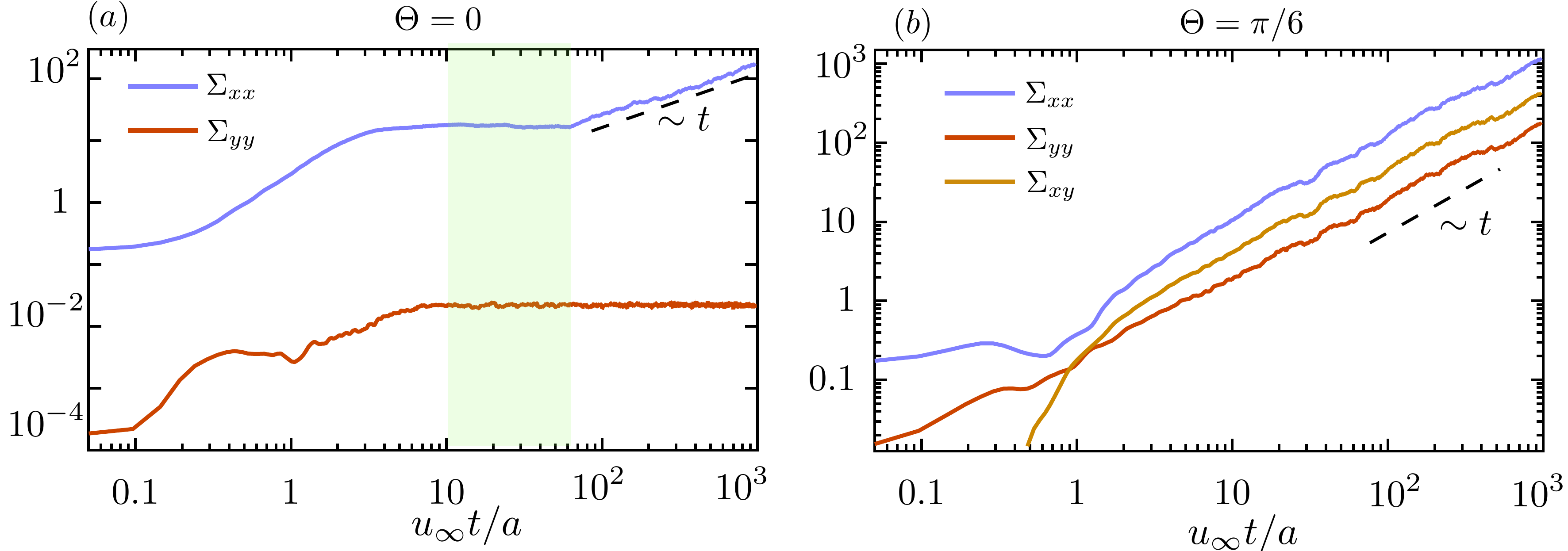}
	\caption{Components of the mean squared displacement tensor as functions of time for: ($a$) $\Theta=0$ and ($b$) $\Theta=\pi/6$. Parameter values are as in Fig.~\ref{fig:angdist}.}
	\label{fig:MSD}
\end{figure}

\begin{figure}[t]
	\centering
	\includegraphics[width=1\linewidth]{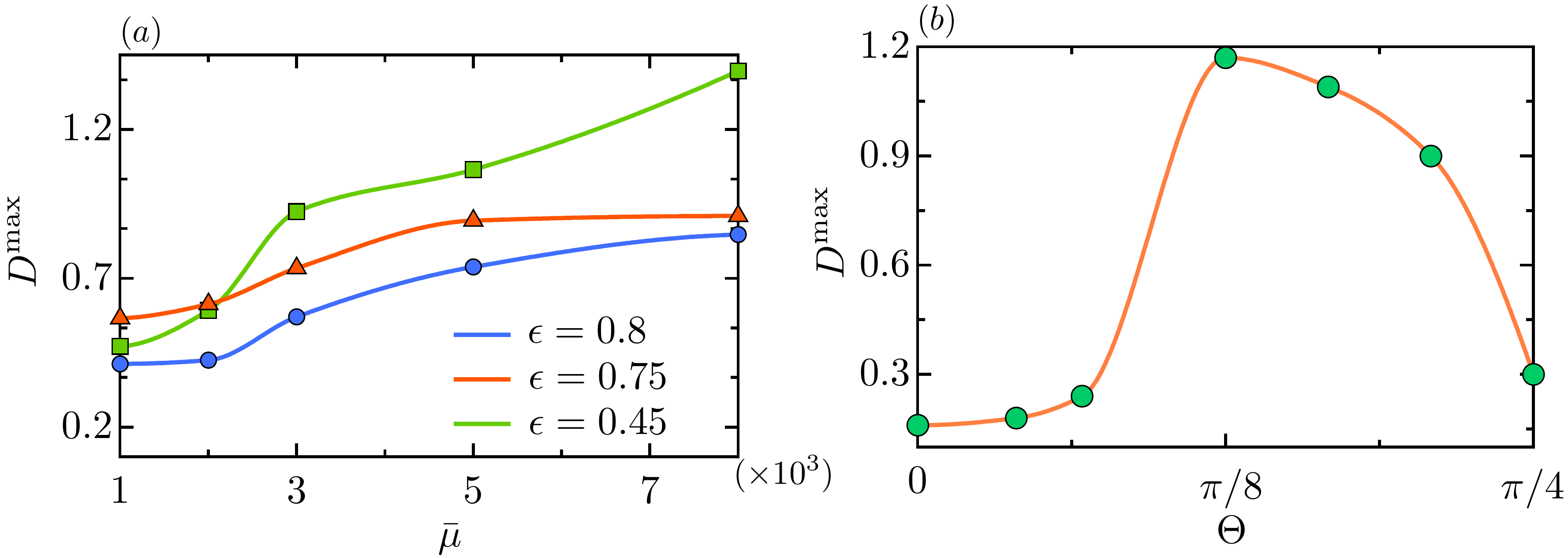}
	\caption{($a$) Variation of $D^{\max}$ with $\bar{\mu}$ at an incidence angle of $\Theta = \pi/6$ for three different lattice porosities. ($b$) Variation of $D^{\max}$ with $\Theta$. \textcolor{black}{Error bars are on the order of the symbol size.} Parameter values are as in Fig.~\ref{fig:angdist}. }
	\label{fig:dmax}
\end{figure}

We now turn to long-time transport properties and focus more specifically on the asymptotic dispersivity, a symmetric tensor defined as
\begin{equation}
\tensbf{D} = \frac{1}{2}\lim_{t \to \infty}\frac{\md}{\md t} \tensbf{\Sigma}(t).
\end{equation}
$\tensbf{\Sigma}(t)$ is the mean-square displacement (MSD) dyadic given by 
\begin{equation}
\tensbf{\Sigma}(t) = \big \langle \left[\bar{\bx}(t) - \langle \bar{\bx}(t)\rangle \right] \left[\bar{\bx}(t) - \langle \bar{\bx}(t)\rangle \right] \big \rangle,
\end{equation}
where $\bar{\bx}(t)$ is the instantaneous filament center-of-mass position and $\langle \cdot \rangle$ denotes the ensemble average. We have computed  $\tensbf{\Sigma}(t)$ for different parameter values by averaging over more than a hundred filament trajectories spanning thousands of unit cells. We first present results for two representative cases with $\Theta=0$ and $\Theta=\pi/6$ in Fig.~\ref{fig:MSD}, allowing to relate the microscopic dynamics to macroscale transport properties. 
Fig.~\ref{fig:MSD}($a$) shows the two relevant components of the MSD as functions of time in the case of $\Theta=0$. Several interesting features stand out. During an initial transient, both $\Sigma_{xx}$ and $\Sigma_{yy}$ start to grow, \textcolor{black}{as the filaments are prepared in different configurations at $t=0$ and thus follow their own initial path. Soon, all the filaments enter nearly periodic trajectories in the squeezing regime as discussed in the previous section: in this regime, they all move ballistically at nearly the same velocity and with hardly any dispersion, leading to a plateau in  $\Sigma_{xx}$ and $\Sigma_{yy}$ (recall that the mean motion is subtracted when calculating the dispersion tensor).} At later times, however, we observe that $\Sigma_{xx}$ starts growing again with a linear time dependence indicative of diffusive transport, while $\Sigma_{yy}$ maintains the same plateau value. The complete saturation of $\Sigma_{yy}$ results from the caged dynamics that restrict the filaments between rows of pillars and strongly hinder any transverse motion other than that due to molecular diffusion as we observed in Fig.~\ref{fig:angdist}($d$). This caging effect, however, is not present in the $x$ direction, and we attribute the linear growth of $\Sigma_{xx}$ at long times to shear-induced Taylor dispersion following the classical mechanism first proposed by Taylor \cite{taylor1953dispersion}.  \textcolor{black}{This mechanism is supported by the observed increase in the dispersivity with flow strength $\bar{\mu}$ in Fig.~\ref{fig:dmax}($a$), though computational limitations prevented us from identifying a clear scaling with $\bar{\mu}$ in strong flows, as one would expect to have for Taylor dispersion.\cite{taylor1953dispersion,brenner1980dispersion,edwards1991dispersion}} We expect that on extremely long time scales molecular diffusion acting in the transverse direction may ultimately result in an increase in $\Sigma_{yy}$ in Fig.~\ref{fig:MSD}($a$), though this regime is not easily captured within the finite duration of our simulations. The situation is quite different for $\Theta=\pi/6$ as shown in Fig.~\ref{fig:MSD}($b$), where all three components specifying $\tensbf{\Sigma}$ are found to grow linearly with $t$ in the asymptotic limit.  The pre-asymptotic time is also shorter in this case as the filament is able to  sample the unit cell efficiently through its chaotic dynamics. 

For an arbitrary incoming flow angle, the dispersion tensor is non-diagonal and can be expressed as $\tensbf{D} = D_1 \mathbf{e}_1\mathbf{e}_1 + D_2 \mathbf{e}_2 \mathbf{e}_2$, where $(D_1, D_2)$ are its eigenvalues with
corresponding eigenvectors ($\mathbf{e}_1$, $\mathbf{e}_2$). In this case, we quantify dispersion by the maximum eigenvalue $D^{\max} = \max(D_1,D_2)$, which dictates the maximum rate of spreading of a dilute cloud of filaments. The dependence of  $D^{\max}$ on flow strength is shown in Fig.~\ref{fig:dmax}($a$) in lattices with varying porosities and for a fixed incidence angle of $\Theta = \pi/6$. In the absence of flow, dispersion is hindered in dense porous media and we expect the effective diffusivity to be minimum for the largest area fraction of pillars (smallest porosity). However, when a flow is applied, shear enhances dispersion \textcolor{black}{by a mechanism similar to} classical Taylor dispersion \cite{brenner2013macrotransport}. In this case, the effect of pillar density is opposite as arrays with large area fractions generate more shear and therefore result in the strongest dispersion.\ This competition between hindrance to transport by trapping and enhancement due to shear leads to a non-monotonic behavior in the dispersivity in \textcolor{black}{weak flows} with respect to porosity as seen in Fig.~\ref{fig:dmax}($a$). \textcolor{black}{In strong flows, it is primarily the shear that dictates the dynamics resulting in a monotonic growth of dispersion with respect to porosity.} It is worth pointing out that dispersion grows monotonically with flow rate for a given porosity, consistent with the macrotransport theory of passive Brownian tracers \cite{brenner1980dispersion}. 

The dependence on incidence angle $\Theta$ for a given geometry and flow strength is illustrated in Fig.~\ref{fig:dmax}($b$).\
The results suggest that dispersion is maximum for $\Theta \approx \pi/8$ and  weakest for $\Theta=0$ and $\pi/4$. This can be appreciated based on the microscopic dynamics discussed above. For a fixed flow strength, a dilute cloud spreads across the lattice most efficiently when the filament trajectories are chaotic, thus promoting rapid separation of nearby polymers. \textcolor{black}{Indeed, as shown in Fig.~\ref{fig:PSD}, the power spectrum of the orientation angle has the slowest decay for $\Theta=\pi/8$, indicative of aperiodic and strongly chaotic dynamics.} Conversely, symmetric flow patterns hinder dispersion due to the quasi-periodic or periodic trajectories that occur in that case \cite{gaspard1995chaotic}.

\section{Chromatographic separation} \label{sec:sec4}

\begin{figure}[b]
	\centering
	\includegraphics[width=1\linewidth]{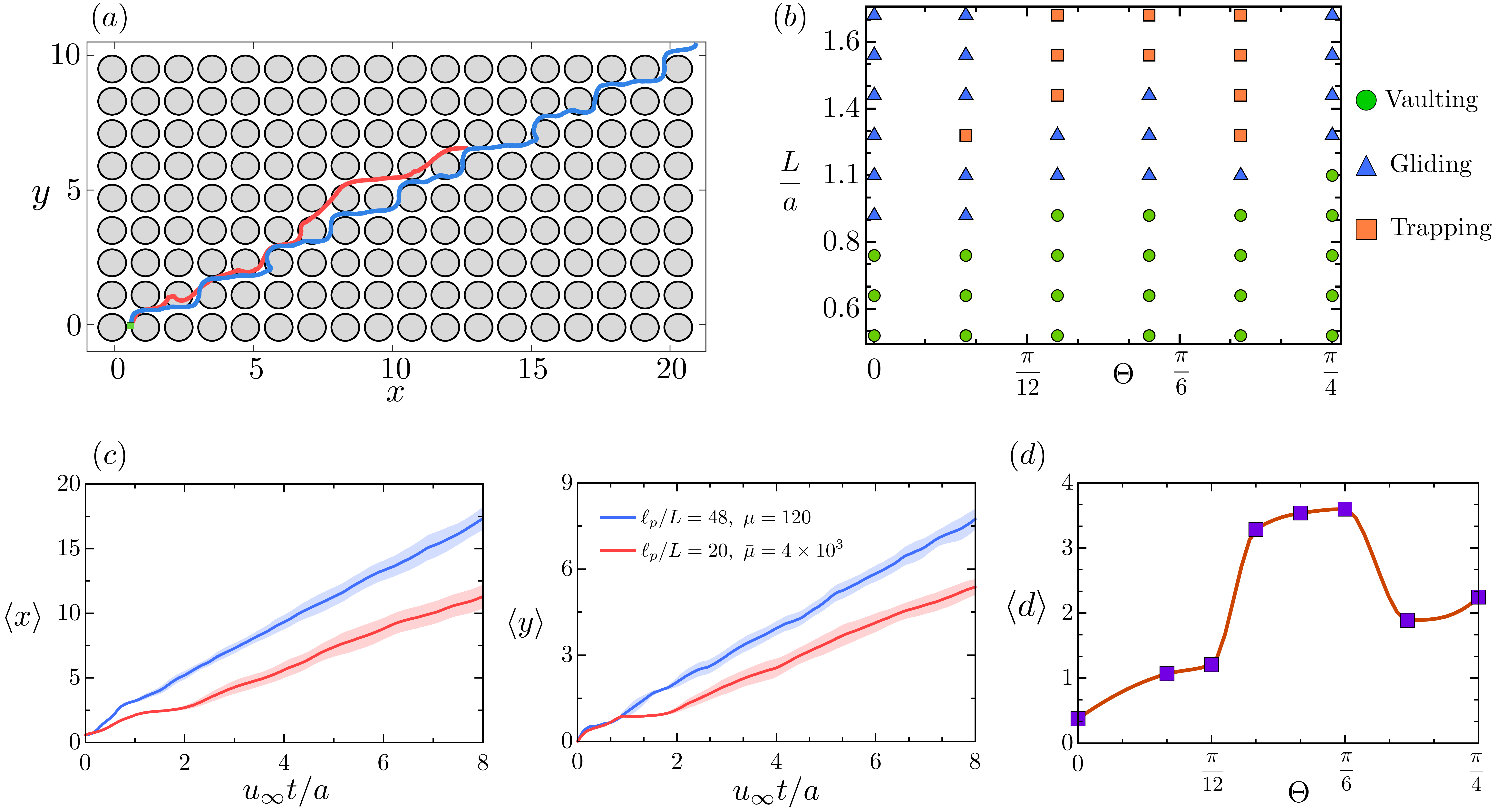}
	\caption{($a$) A typical center-of-mass trajectory of two polymers of lengths $L/a = 0.7$ (blue) and $L/a = 1.7$ (red) over a given period of time with flow at $\Theta = \pi/6$. Both polymers started from the marked point at $t=0$. \textcolor{black}{The elastoviscous numbers are $\bar{\mu} = 120$ and $4 \times 10^3$ for the shorter and longer filaments respectively.} ($b$) Phase chart showing the dominant modes of transport as a function of incidence angle $\Theta$ and contour length $L/a$. ($c$) Ensemble averaged $x$ and $y$ coordinates of the COM of the two polymers for the case shown in (a). ($d$) Time-averaged separation of the two polymers as a function of incidence angle. In all the simulations, $\lambda = 1.2$.}
	\label{fig:sort}
\end{figure}

Not only do scattering dynamics control long-time dispersion, but they also dictate the mean transport velocity. We now discuss how this effect can be leveraged for the use of 2D micro-patterned porous media as  chromatographic devices that can sort polymers based on their lengths.

As an illustrative example, we consider two filaments of lengths $L/a = 0.7$ and $L/a = 1.7$. Assuming the two polymers have the same persistence length $\ell_p$, the shorter filament is effectively experiencing weaker thermal shape fluctuations. In a typical microfluidic experiment, a macroscopic pressure gradient drives the flow through the porous medium, which sets the characteristic shear rate $u_\infty/a$ of the problem. As a result, the longer polymer has a larger elastoviscous number that scales as $\sim L^4$. Fig.~\ref{fig:sort}($a$) shows typical COM trajectories of the two polymers over a given period of time, starting from the same position highlighted on the figure at $t=0$. It is evident from the Lagrangian trajectories that over the course of time the filaments separate out quite efficiently after interacting only with $\sim$ 6--7 pillars. The shorter polymer (shown in blue) experiences an elastoviscous number that is below the buckling threshold \cite{liu2018morphological}. As a result, it slides past the obstacles and relies on the \emph{vaulting} mode discussed in Section \ref{sec:sec2} to get transported without much effective hindrance. On the other hand, the longer polymer (shown in red) frequently gets trapped around the obstacles and mainly adopts folded conformations due to buckling instabilities. Such wrapping of the polymer around the obstacles is made evident by the frequent intersections between the COM trajectory and the pillars in Fig.~\ref{fig:sort}($a$). Ensemble averaged trajectories of the COM coordinates are plotted as functions of time in Fig.~\ref{fig:sort}($c$) and clearly show the increasing separation of the two polymers through their interaction with the pillars and the flow.

We quantify the efficiency of this separation process using the mean separation distance defined as:
\begin{equation}
\langle d \rangle(T) = \frac{1}{T}\int_0^T |\bar{\bx}_1(t) - \bar{\bx}_2(t)|\,dt.
\end{equation}
Fig.~\ref{fig:sort}($d$) shows the dependence of $\langle d \rangle(T)$ on the flow incidence angle $\Theta$ for a dimensionless time of $T=8$. Consistent with the results on dispersivity shown in Fig.~\ref{fig:dmax}($b$), the separation $\langle d \rangle(T)$ is maximized for an incidence angle close to $\Theta = \pi/6$, which we attribute to the chaotic scattering dynamics of the Lagrangian trajectories. We also find as expected that symmetric flow patterns at $\Theta = 0$ or $\Theta = \pi/4$ also result in poor separation; see the supporting movies for a visual illustration of these differences.

It is evident that in the process of such a numerical design it is important for the two polymers under consideration to exhibit distinct scattering dynamics. In order to appreciate this, we present a phase chart in Fig.~\ref{fig:sort}($b$) showing the dominant  mode of transport as a function of the flow angle $\Theta$ and contour length $L/a$.  \textcolor{black}{In order to classify the different modes of transport in a systematic way, we first studied the end-to-end distance $R_{ee}$ of the polymers. If $R_{ee}$ remained within 12\% of the contour length of the polymer, the mode was classified as vaulting. Otherwise, when larger deformations occured, we then considered the velocity of the center-of-mass. The trapping mode was identified by detecting the presence of prolonged plateaus of the velocity near the value of zero. Note, however, that typically the trapping mode alternates with events of squeezing or gliding. In order to distinguish them, we looked at the dominant events during the interaction of the polymer with one hundred pillars.}
The phase chart eludes to the fact that the optimal angle $\Theta$ for separation is expected to be different depending on the length of the polymers that one wants to separate.

\section{Concluding remarks}\label{sec:sec5}

\begin{figure}[b]
	\centering
	\includegraphics[width=1\linewidth]{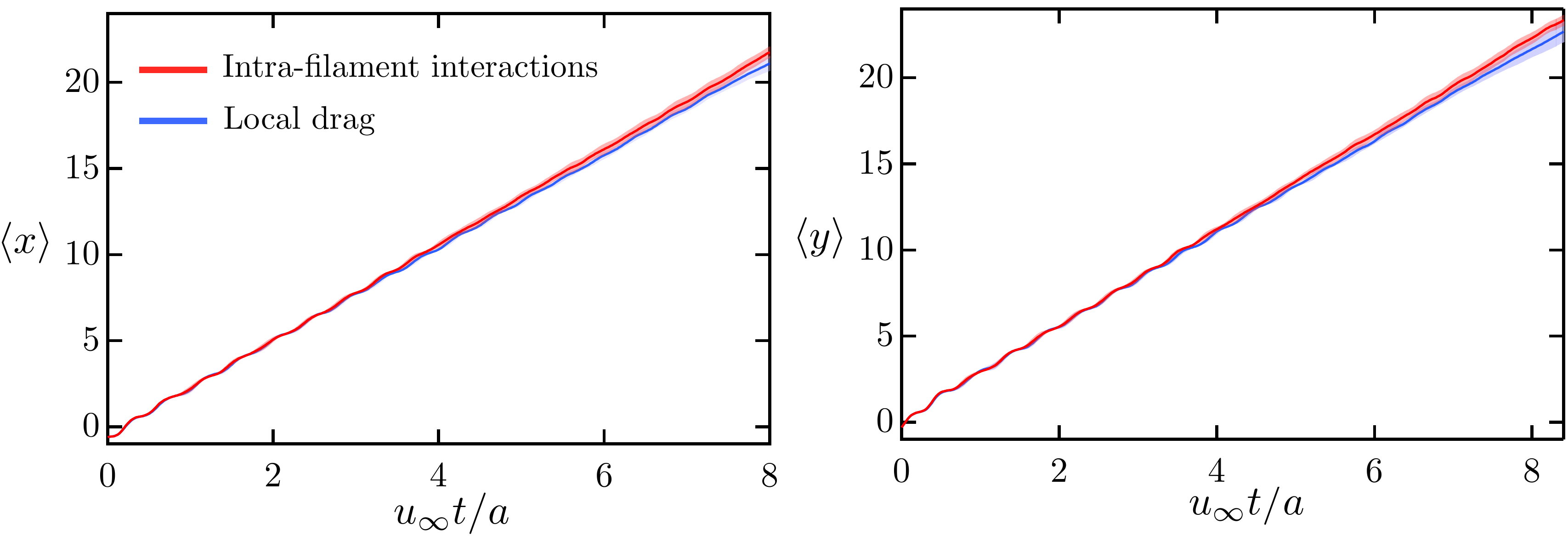}
	\caption{\textcolor{black}{Ensemble averaged $x$ and $y$ coordinates of the center of mass as functions of time, comparing results from the local drag model with a model including intrafilament hydrodynamic interactions. Parameter values: $L/a=1$, $\Theta = \pi/4$ and  $\bar{\mu} = 8 \times 10^3$.}}
	\label{fig:locHI}
\end{figure}

We have analyzed the long-time transport properties of semiflexible polymers with $L \sim \ell_p$ in structured porous media under the action of an imposed flow. In contrast to entropic polymer molecules, the dynamics in this case is governed  by a competition between dynamic buckling instabilities  and interactions with obstacles, which depend on several geometric factors such as  the incidence angle of the imposed flow, the lattice porosity and polymer contour length. These complex interactions were found to result in three dominant modes of transport, coined trapping, gliding and vaulting. 
In the spirit of recently studied bacterial spreading in microfluidic crystals \cite{dehkharghani2019bacterial}, our simulations revealed how these scattering dynamics and transport in the non-uniform flow result in long-time diffusive transport. Following the classical mechanism of Taylor dispersion \cite{taylor1953dispersion}, shear is typically found to enhance dispersion, yet strong flows also lead to more frequent trapping of polymers around obstacles, a phenomenon unique to deformable particles that tends to reduce transport.

Leveraging these scattering dynamics, we demonstrated how an array of microposts can be used to sort polymers according to their contour length. This chromatographic design bears similarities with the post arrays often used for the separation of long- chain polymers \cite{patel2003computational,mohan2007stochastic,dorfman2012beyond}, as well as with deterministic lateral displacement devices (DLD) that are classically used to sort biological cells based on their size and deformability \cite{karabacak2014microfluidic,o2019cooperative}.  The explored parameter space reveals how the angle of the flow can be optimized for efficient sorting of the filaments. Further optimization work, however, would be useful, in particular to identify the role of pillar shape on dispersion and separation \cite{kabacaouglu2018optimal}.

\textcolor{black}{All the numerical results presented here relied on a local anisotropic drag model for the filament dynamics, which neglects the role of hydrodynamic interactions in the spirit of previous studies involving DNA molecules \cite{kim2007brownian,teclemariam2007dynamics,cho2010brownian}. To assess the role of intrafilament hydrodynamic interactions, we performed an additional set of simulations retaining the non-local operator in the slender-body equation \cite{tornberg2004simulating,keller1976slender}.   Fig.~\ref{fig:locHI} compares the center of mass trajectories over approximately 30 pillars for both models, where it is evident that the effect of intrafilament hydrodynamic interactions is weak. Our model also glosses over  polymer-pillar hydrodynamic interactions, which are more challenging to capture numerically. Andr\'e \emph{et al.} \cite{andre1998polyelectrolyte} showed that these  interactions are not significant during the mechanical hooking and unhooking of DNA molecules past micro-pillars during electrophoretic transport, and that local drag models can provide quantitative results. Including hydrodynamic interactions with pillars would require accounting for lubrication films that arise during close contacts of the polymers with  pillars, for instance during trapping events. We speculate that these lubrication films will slow down the approach of the polymers towards the pillars and also prolong their escape as tangential motion will incur additional viscous dissipation, and this may possibly lead to a reduction in the mean velocity and dispersivity. Nevertheless, we believe that our leading-order hydrodynamic model still provides useful qualitative insight in the scattering dynamics and chromatographic separation  process. Future work will carefully address the possible role of the polymer-pillar hydrodynamic interactions and extend the present results to non-dilute polymer solutions. }

\vspace{0.3cm}
D.S. gratefully acknowledges funding from NSF Grant CBET-1934199. 

\appendix

\section*{Appendix: Contact algorithm}\label{app:contact}

We outline the algorithm used to prevent penetration of the filaments into the pillars, which is inspired by the work of Evans \textit{et al.}\cite{evans2013elastocapillary}. The local SBT equation (\ref{eq:sbt}) can be re-arranged as follows:
\begin{equation}
\frac{\partial \bx}{\partial t} + \tensbf{\Lambda} \cdot \bx_{ssss} =\bar{\mu} \bu - \tensbf{\Lambda} \cdot \left(-(\sigma \bx_s)_s + \sqrt{L_f/\ell_p} \boldsymbol{\xi}\right) = \mathbf{F},
\end{equation}
where $\mathbf{F}$ contains terms due to the background flow, internal tension, and Brownian forces. For any given point $\bx(s,t)$ along the filament,  we first identify the cell center $\bx_c =  (x_c,y_c)$ in which it is located. Let $d$ denote the Euclidean distance between $\bx(s,t)$ and that cell center. Following Evans \textit{et al.}\cite{evans2013elastocapillary}, we introduce the unit vector $\mathbf{\hat{p}}$ defined as:

\begin{equation}
\mathbf{\hat{p}} = \frac{\bx(s,t)-\bx_c}{d}.
\end{equation}
If $d-a < \varepsilon$, where $\varepsilon$ is a small cut-off distance, we project the force $\mathbf{F}$ parallel and perpendicular to $\mathbf{\hat{p}}$ to define:
\begin{align}
F_n &= \mathbf{\hat{p}} \cdot \mathbf{F}, \\
\mathbf{F}_t &=\left(\mathbf{I}-\mathbf{\hat{p}} \mathbf{\hat{p}}\right) \cdot \mathbf{F}.,
\end{align}
where subscripts $n$ and $t$  stand for the normal and tangential directions, respectively. Keeping $\mathbf{F}_t$ unchanged, we alter the component $F_n$ as follows \cite{evans2013elastocapillary}:
\begin{equation}
F_n = \mathrm{min}\left[F_n,  \left(1-\left(\frac{\varepsilon}{d-a}\right)^m\right)F_n \right],
\end{equation}
where we chose $m=4$. The above definition is such that if the fluid, tension and Brownian forces try to separate the filament from the pillar then the force component is unaltered. However, if these same forces are attempting to push the filament into a pillar then the sign of the force is reversed. Since the tangential component is unchanged, this treatment allows the filament to glide or wrap around the obstacles without significant numerical difficulties. In the event of overlap with a pillar, we adaptively reduce the time step to ensure numerical stability. In all the simulations shown here, we have used $\varepsilon = 0.005 L $.
\begin{figure}[H]
	\centering
	\includegraphics[width=0.35\linewidth]{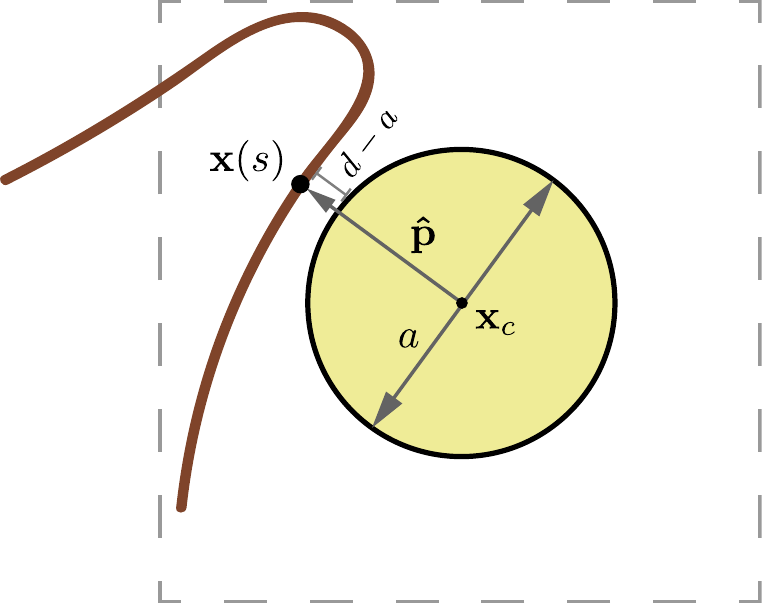}\vspace{-0.cm}
	\caption{Schematic of a polymer close to contact with a pillar and relevant variables.} 
\end{figure}

\bibliography{rsc} 
\end{document}